\documentclass[english]{article}
\usepackage[T1]{fontenc}
\usepackage[latin9]{inputenc}
\usepackage{geometry}
\usepackage{color}
\usepackage{graphicx}
\usepackage{subcaption}
\usepackage{amssymb}
\usepackage{epstopdf}
\usepackage{amsmath}
\usepackage{caption}
\newcommand{\comments}[1]{}   

\makeatletter

\usepackage{indentfirst}
\usepackage{enumerate}

\makeatother

\begin{document}

\title{\bf Role of initial conditions in the dynamics of \\
quantum glassy systems}
\vskip 10pt
\author{
Leticia F. Cugliandolo$^1$, Gustavo S. Lozano$^2$ and Nicol\'as Nessi$^2$\\
$\; ^1$Sorbonne Universit\'e,
Laboratoire de Physique Th\'eorique et Hautes Energies, UMR 7589 CNRS,\\
Tour 13, 5\`eme \'etage,  4 Place Jussieu, 75252 Paris Cedex 05, France\\
$\; ^2$Departamento de F\'{\i}sica,
FCEYN Universidad de Buenos Aires \& IFIBA CONICET,\\
Pabell\'on 1 Ciudad Universitaria, 1428 Buenos Aires, Argentina
}
\date\today
\maketitle

\begin{abstract}
	We set the formalism to study the way in which the choice of canonical equilibrium initial
conditions affect the real-time dynamics of quantum disordered models.
We use a path integral formulation on a time contour with real and
imaginary time branches. The
factorisation of the time-integration paths usually assumed in field-theoretical studies breaks
down due to the averaging over quenched randomness.
We derive  the set of Schwinger-Dyson dynamical
equations that govern the evolution of linear response and correlation
functions. The solution of these equations
is not straightforward as it needs, as an input,
the full imaginary-time (or Matsubara frequency) dependence of the
correlation in equilibrium. We check some limiting cases (equilibrium dynamics, classical limit) and
we set the stage for the analytic and numerical analysis of quenches in random quantum systems. 
\end{abstract}

\section{Introduction}

The aim of this paper is to introduce the formalism that allows one to study the real-time dynamics of
interacting quantum systems with quenched randomness starting from equilibrium represented
by a canonical density operator and evolving  under generic conditions that could include a
so-called quantum quench~\cite{Bloch08,Polkovnikov10,Gogolin,Pasquale-ed}. The quenched randomness is modelled in the form of frozen coupling strengths
taken from a probability distribution or, more generally, a random potential. For concreteness, we
 focus on models of mean-field type for which a complete description is given in terms of
 functions of two times only. The relevant dynamic equations that couple these two-time dependent
 objects are derived using the Schwinger-Keldysh closed-time contour method in its functional integral formulation~\cite{ZhSuHaLu85,GrScIn88,vanLeeuwen06,Kamenev11,Stefanucci}.
 They take the form of Schwinger-Dyson integro-differential
 equations that couple in a highly non-trivial way the correlation and
 linear response functions with support on the closed time path contour in the complex plane.
After a series of steps that we describe in detail in Sec.~\ref{sec:path-integral} we write these equations using a relatively compact
 notation that condenses all non-linearities in the self-energy $\Sigma$ and vertex $D$ functions that depend themselves on the
 various correlations and linear responses. Since the model has already been studied in detail in full equilibrium and
 after a quench from infinite temperature  
 (that is to say, from an initial state that is not correlated with disorder) under dissipative conditions ({\it i.e.}, coupled to a quantum bath)
 we can
 take different limits and confirm that our generalised equations take the expected forms.

An important feature of our construction is that the full set of dynamic equations involve correlations evaluated at real times, at purely imaginary times (on the Matsubara branch of the contour) as well as mixed ones with real and imaginary time dependences. This is different from what was advocated in studies of thermal field theories without randomness, see {\it e.g.}~\cite{Niemi1,Niegawa99}, were it was claimed that the imaginary and real time brach contributions simply factorise. The importance of keeping these mixed correlations was reckoned in the condensed matter literature and it is, indeed, discussed in~\cite{vanLeeuwen06,Stefanucci}.

The formalism here introduced will be of use to study questions as the thermalisation (or not) of 
quantum disordered systems after sudden changes in some parameters are the interplay between 
glassy properties and many-body localisation. Indeed, the random potential that we focus on here,
as an example, has been used  to analyse many-body localisation~\cite{NaHu15,AlVo15} issues 
at the mean-field level in,
{\it e.g.}~\cite{Baldwin16,Baldwin17}. The complete solution of the dynamic equations, starting 
from thermal initial states with different characteristic (of the disordered phase, in a metastable 
situation, in the ordered glassy phase) should lead to a dynamic phase diagram as the one 
discussed in~\cite{CuLoNe17}, now extended to the quantum problem.

\section{The model}
\label{sed:model}

For concreteness, we consider the quantum extension of the classical
spherical $p$ spin model~\cite{CrSo92,CuKu93,Ba97}, for which a ``kinetic energy''
term is added to the usual potential energy~\cite{CuLo98,CuLo99,CuGrSa00,CuGrSa01,BiCu01,CuGrLoLoSa02,Rokni04,Gracia04}. The
choice of this model is just based on the fact that it has been studied in
great detail in equilibrium and out of equilibrium situations that serve as
different controlling limits of our development. Having said this, the
method that we use is very general and the structure of the equations that we
derive will be similar for a generic field theory with a quenched random
potential.

\subsection{Definition}

We represent the $N$-dimensional ``coordinate'' by $\hat {\bf s}\equiv (\hat s_1,......, \hat s_N)$
and the $N$-dimensional ``momentum'' by
$\hat {\bf \pi}\equiv (\hat \pi_1,......, \hat \pi_N)$. We then quantify the model by upgrading them
to be canonically conjugated
\begin{equation}
[\hat \pi_i, \hat s_j ] = {\rm i} \hbar \delta_{ij}
\end{equation}

The kinetic energy is~\cite{CuLo98,CuLo99}
\begin{eqnarray}
\hat H_{\rm kin}[\{\hat \pi_i\}] &=& \frac{1}{2m} \sum_{i=1}^N \hat\pi^2_i
\; ,
\label{eq:pspin-kin}
\end{eqnarray}
The parameter $m$ is the mass of the particle.

The $p$-spin model is a model with interactions between $p$ spins mediated by
quenched random couplings $J_{i_1 \dots i_p}$.
The quantum potential energy is the natural extension of the classical one used
in~\cite{KiTh87a,KiTh87b,KiWo87,De80,GrMe84} in which the
coordinates are now position-like operators
\begin{eqnarray}
\hat H_{\rm pot}[\{\hat s_i\}] =
- \frac{1}{p!} \sum^N_{i_1 \leq \dots \leq i_p} J_{i_1 \dots i_p} \hat s_{i_1} \dots \hat s_{i_p}
\; .
\label{eq:pspin-pot}
\end{eqnarray}
The coupling exchanges are independent identically distributed random variables taken from a Gaussian distribution with average and variance
\begin{equation}
[ J_{i_1 \dots i_p } ] =0
\; , \qquad\qquad
[ J^2_{i_1 \neq \dots \neq i_p } ] = \frac{J^2 p!}{2N^{p-1}}
\; .
\label{eq:disorder-statistics}
\end{equation}
The parameter $J$ characterises the width of the Gaussian and 
$p$ is an integer larger than or equal to $2$ that defines the model.
In the large $N$ limit we (naively) impose a spherical constraint on the operator $\hat {\bf s}$ by adding the term
\begin{equation}
\hat H_{\rm constr} = \frac{z}{2} \ \left(\sum \hat s_i^2 - N \right)
\end{equation}
to the Hamiltonian, with $z$ a Lagrange multiplier.

The  {\it Hamiltonian (spherical) $p$-spin model}  is then defined by
\begin{equation}
\hat H_{\rm syst} = \hat H_{\rm kin} + \hat H_{\rm pot} + \hat H_{\rm constr}
\; .
\label{eq:pspin-total-energy}
\end{equation}
This model represents a quantum particle constrained to move on an $N$-dimensional hyper-sphere
with radius $\sqrt{N}$ in the large $N$ limit.

The potential energy (\ref{eq:pspin-pot}) is one instance of a generic random potential $V(\{\hat s_i\})$
with zero mean and correlations~\cite{En93,FrMe94,CuLe96}
\begin{equation}
[ V(\{\hat s_i\})  V(\{\hat s'_i\}) ] = -N {\cal V}(|\hat {\bf s}-\hat{\bf s}'|/N)
\end{equation}
with $ {\cal V}(|\hat{\bf s}-\hat{\bf s}'|/N) =-  \frac{J^2}{2} (\hat{\bf s} \cdot \hat{\bf s}'/N)^p$.
The problem is also interesting
for generic ${\cal V}$ but we will focus here on the
monomial case that corresponds to the $p$-spin model. The changes induced by other functions
${\cal V}$ are easy to derive.

This model can be interpreted in several ways. As previously mentioned,
it is a possible quantum extension of the spherical $p$ spin model.
It also describes a non-linear generalisation of the
quantum rotor spin-glass models discussed in the literature~\cite{YeSaRe93}.
Finally, its partition function is formally identical to that of a classical
polymer of length $L=\beta\hbar$ in an $N$ dimensional space.

We will be interested in studying the dynamics using canonical equilibrium initial
conditions. These will be imposed by using a density operator at the
initial time, $t=0$, of the Gibbs-Boltzmann form
\begin{equation}
\hat \rho(0) = {\mathcal Z}_{\rm eq}(\beta) \ e^{-\beta \hat H_0}
\; .
\end{equation}
The Hamiltonian $\hat H_0$ depends explicitly on the
quenched randomness. This dependence entails a technical complication
due to the fact that the average over randomness needs the use of the replica
method, differently from what happens for initial conditions that are uncorrelated
with the fixed randomness~\cite{CuLo98,CuLo99}. We explain how this averaging
procedure is done in Sec.~\ref{sec:path-integral}.

The initial Hamiltonian and the one used in the time evolution are both given in Eq.~(\ref{eq:pspin-total-energy}). We
keep them different to allow for the study of instantaneous quenches. At a given point in the presentation we
will specify the quench to be
\begin{equation}
J^0_{i_1\dots i_p} \mapsto J_{i_1\dots i_p}  = \frac{J}{J_0} \ J^0_{i_1\dots i_p}
\; ,
\end{equation}
operated at time $t=0$, 
and this choice will allow us to perform the average over disorder.

\subsection{Background}

The mean-field character of the model allows one to attack it  from different angles with various analytic methods
and hence deduce its equilibrium, metastable and out of equilibrium properties. 
Let us describe below, in a few words, what is 
known about the cases with $p\geq 3$.

The study of the canonical Boltzmann equilibrium properties of the $p>2$ quantum model using the replica technique was
presented in~ Refs.~\cite{CuGrSa00,CuGrSa01}. The equilibrium phase diagram was derived in these
papers. Its peculiarity is that the transition line separating the disordered and the ordered phases
is of second order (though discontinuous with respect to the order parameter) close to the classical critical point and changes to  first order close to the
quantum critical point at zero temperature. This feature is common in quantum mean-field disordered models
with a one-step replica symmetry breaking solution in the ordered phase. It has implications in the
failure of quantum annealing algorithms in hard optimisation problems, such as the $K$-sat
model with $K>2$~\cite{Bapst}. The disordered phase is of paramagnetic kind but there are two different
such states: one is dubbed a classical paramagnet and the other one a quantum paramagnet. The former exists at
high temperatures and weak quantum fluctuations, while the latter is the equilibrium phase in the opposite limit
of low temperatures and strong quantum fluctuations. As in the strict classical limit, close to the second
order phase transition, in a finite region of the phase diagram the classical paramagnet is not a simple one
but a mixture of an exponentially large number of metastable (TAP) states.

Interestingly enough, the so-called static approximation~\cite{BrMo} that amounts to using a constant
(imaginary time independent) order parameter $q_d(\tau) = q_d$ and $q_d$ a variational
parameter gives correct qualitative results. It predicts the topology of the phase diagram. This
approximation should also be useful in a first analysis of the asymptotic limit dynamics, needing
less numerical computational effort than the full solution of the coupled set of differential equations
that we will derive.

The analysis of the quantum Thouless-Anderson-Palmer (TAP) equations was presented in~\cite{BiCu01}. It yields a
way to derive the free energy landscape of the quantum model and this, as in the classical
case, allows one to interpret and better understand the static and dynamic behaviour.
In particular, this method shows which is the character of the classical paramagnet, as a
mixture of metastable states, close to the transition line.

The Schwinger-Keldysh generating function~\cite{Sc61,Ke64,Ke65} for the real-time dynamics
of this model coupled to a quantum bath
was obtained in~\cite{CuLo98,CuLo99} (see~\cite{ZhSuHaLu85,GrScIn88,vanLeeuwen06,Kamenev11,Stefanucci} for reviews of this
technique). From it, the real-time dynamic equations starting from random initial conditions, that one
can associate to an infinite temperature initial state
and model with  a constant density operator,  were
derived. The case of the model being infinitesimally coupled to an Ohmic bath was considered in~\cite{CuLo98,CuLo99}
and stronger couplings to Ohmic and more generic non-Ohmic baths were dealt with in~\cite{CuGrLoLoSa02}.
Their asymptotic behaviour was obtained with approximate analytic methods and the numerical
data stemming from the full solution of the equations confirmed these results.

Similar studies of other mean-field disordered models were conducted in the 90s and until recently. 

The formalism that we develop in the next Section will allow one to study the dynamics 
of the isolated system following quenches from initial thermal states, including the possibility 
of setting the system in a TAP state initially and studying whether it remains trapped in it or 
escapes from it, in the manner presented in~\cite{CuLoNe17,CuLoNePiTa18} for the 
classical counterparts with $p>2$ and $p=2$, respectively.

\section{Closed-time path-integral formalism}
\label{sec:path-integral}

The approach that we use to derive the dynamic equations results from the combination
of different techniques that have been discussed, in different contexts, in the
literature. Essentially, it consists of the Schwinger and Keldysh~\cite{Sc61,Ke64,Ke65} closed-time path-integral formalism
(for reviews, see~\cite{ZhSuHaLu85,GrScIn88,vanLeeuwen06,Kamenev11,Stefanucci}) with the addition of a method developed to impose
equilibrium initial conditions in systems with quenched randomness~\cite{HoJaYo83}
that we adapt here to quantum problems.

\subsection{Operators}

For the moment we do not specify the Hamiltonians that we will use to build the initial conditions 
nor the further time evolution. We simply define, in this Section, a number of correlation functions 
that will be useful in the development of the method. The Hamiltonian $\hat H$ used in the following 
expressions should therefore be taken to be a generic one.

We define the evolution operator
\begin{equation}
\hat U(t,t_0)= \hat T \exp\left[-{\rm i} \int_{t_o}^t dt \, \hat H(t)\right]
\end{equation}
with $\hat T$ the time ordering.
We use the Heisenberg picture in which operators depend on time according to
\begin{equation}
\hat s_i(t) = \hat U(t_0,t) \, \hat s_i(t_0) \,  \hat U(t,t_0)
\; .
\end{equation}
In what follows the Hamiltonians will be time-independent and
only abrupt changes will be considered in which it changes instantaneously from
one form $\hat H_0$ at time $t_0=0$ to $\hat H$ at $t>0$. Strictly speaking, 
no time integral nor time ordering are needed in the evolution operator. Still, 
we may introduce a formal time dependence in $\hat H$ for $t>0$ in order
to define response functions.

The {\it symmetric} real time dependent correlation functions are defined as
\begin{equation}
C_{ij}(t,t^{Ž})=\frac{1}{2}\langle (\hat s_i(t) \hat s_j(t') +\hat s_j(t') \hat s_i(t) )\rangle = 
\frac{1}{2}\mbox{Tr} \left[ \,( \hat s_i(t) \hat s_j(t')+\hat s_j(t') \hat s_i(t) )\, \hat \rho(0) \right]
\end{equation}
where $\hat \rho(0)$ is the initial density operator.
In terms of the lesser $G_{ij}^<(t,t')=-{\rm i}\langle \hat s_i(t') \hat s_j(t) )\rangle$ and greater 
$G_{ij}^>(t,t')=-{\rm i}\langle \hat s_j(t) \hat s_j(t') )\rangle$ Green functions, the symmetric correlation
function reads
\begin{equation}
C_{ij}(t,t')=- {\rm i} [G_{ij}^>(t,t')+G_{ij}^<(t,t')]
\; . 
\end{equation}
Times $t$ and $t'$ appear in one and the reversed order in the two terms contributing to the symmetric correlation. 
The way to adapt a time-ordered calculation to yield this not  time ordered correlation function is to double the
real time path letting it go from the initial time 0 to a final time $t_f$ and then come back from
$t_f$ to 0. These two branches, with time running forwards and backwards, respectively,
are then called $+$ and $-$. We will introduce them explicitly in the construction of the closed-time path integral.

The initial state is characterised by the operator density $\hat \rho(0)$. As already said, we will choose to work with 
thermal states, such that 
\begin{equation}
\hat \rho(0)=\frac{e^{-\beta \hat H_0}}{{\mathcal Z}_{\rm eq}(\beta)} 
\qquad\mbox{with} \qquad {\mathcal Z}_{\rm eq}(\beta) =\mbox{Tr} e^{-\beta \hat H_0}
\end{equation}
where $\beta= 1/(k_BT)$ is the inverse temperature.
In order to take into account correlations in the initial state,  an imaginary time leg can be attached to the closed-time 
path, with the {\em imaginary time parameter} $\tau \in  [0,- {\rm i} \beta]$.

As it is standard in this field, together with the symmetrised correlation function we can define the (advanced) linear 
response function
\begin{equation}
R_{ij}(t,t')= \frac{{\rm i}}{\hbar} \theta(t-t')\langle [\hat s_i(t), \hat s_j(t')] \rangle 
\; , 
\end{equation}
here expressed as a commutator between the operators $\hat s_i(t)$ and $\hat s_j(t')$ thanks to the 
Kubo relation valid in and out of equilibrium.

Together with these real time correlation functions, we introduce the Matsubara Green function,
\begin{equation}
q_{ij}(\tau, \tau')= \langle \hat T_\tau \hat s_i(\tau) \hat s_j(\tau') \rangle =    
\mbox{Tr} \left[ \, \hat T_\tau \hat  s_i(\tau) \hat s_j(\tau') \, \hat \rho(0) \right]
\end{equation}
(we recall that $\tau \in (0, -{\rm i} \beta\hbar)$), 
where $ \hat T_\tau$ is the time ordering operator along the imaginary time axis
\begin{equation}
\hat s_i(\tau)= \hat U_0 (0,-{\rm i} \tau) \hat s_i(0) \hat U_0 (-{\rm i} \tau,0) 
\qquad\mbox{with}\qquad 
\hat U_0 (-{\rm i} \tau,0)=e^{-\tau \hat H_0}
\; . 
\end{equation}
The Matsubara Green function encodes the information of correlations in the initial state.
In the same way, it will be convenient to define the correlations with mixed time arguments, i.e.  with one argument in  real time and the other in the imaginary sector by
\begin{equation}
G_{ij}^\rceil(t,\tau)=-{\rm i}\langle \hat s_i(t) \hat s_j(\tau) \rangle =  
 -{\rm i}\mbox{Tr} \left[ \, \hat s_i(t) \hat s_j(\tau) \, \hat \rho(0) \right]
\end{equation}
\begin{equation}
 G_{ij}^\lceil(\tau, t)=-{\rm i}\langle \hat s_i(\tau) \hat s_j(t) \rangle = 
 -{\rm i}  \mbox{Tr} \left[ \, \hat s_i(\tau) \hat s_j(t) \, \hat \rho(0) \right] 
 \; . 
 \end{equation}
In this case no time ordering is needed since by convention, imaginary times occur later than any real time in our contour.

\subsection{The path-integral}

The ``in-in'' generating functional is defined as
\begin{eqnarray}
{\mathcal Z}[{\bf y}^+,{\bf y}^{-}] =
\mbox{Tr} \left[ {\hat T}^*\exp\left(-{{\rm i}\over \hbar}
\int_0^{t_f} dt\,\, {\bf y}^{-}(t)\cdot \hat {\bf s}(t)\right)
\; \hat T \exp\left({{\rm i}\over \hbar}
\int_0^{t_f} dt\,\, {\bf y}^{+}(t)\cdot \hat {\bf s}(t)\right)
\hat{\rho} (0)\right] ,
\label{gf}
\end{eqnarray}
where ${\bf y}^+$ and ${\bf y}^{-}$ are $N$-vector external
sources, $\hat{\rho} (0)$ is the density operator at the initial
time, $t_0=0$, and the symbols T and T$^*$ are the time and
anti-time ordering operators, respectively.
${\mathcal Z}[{\bf y}^+,{\bf y}^{-}]$
admits the following path-integral representation~\cite{Kamenev11}:
\begin{eqnarray}
{\mathcal Z}[{\bf y}^+,{\bf y}^{-}]= \int D{\bf s}^+ D{\bf s}^-
\;
e^{{{\rm i}\over \hbar}\left(S[{\bf s}^+]-S[{\bf s}^{-}]+
\int dt\,\,
{\bf y}^{+}(t)\cdot {\bf s}^+(t)
-\int dt\,\,
 {\bf y}^{-}(t)\cdot {\bf s}^-(t)\right)
 }
\; \langle{\bf s}^+|\hat{\rho}(0)|{\bf s}^-\rangle
\; ,
\label{gf1}
\end{eqnarray}
where $\langle {\bf s}^+|\hat{\rho}(0)|{\bf s}^-\rangle \equiv
\langle {\bf s}^+(0)|\hat{\rho}(0)|{\bf s}^-(0) \rangle$ stands for the matrix element
of the density matrix with support  at $t=0$ only. The doubling of
degrees of freedom, $({\bf s}^+,{\bf s}^-)$, is a consequence of
having to slice two real time branches, one running forwards and the other one running backwards
to compute the two matrix elements,
one for each source $({\bf y}^+,{\bf y}^{-})$. 
The integrals over paths are short-hand notations for
\begin{equation}
 \int D{\bf s}^+ D{\bf s}^- \dots =
  \int_{{\bf s}^+(0)}^{{\bf s}^+(t_f)} D{\bf s}^+ \int_{{\bf s}^-(t_f)}^{{\bf s}^-(0)} D{\bf s}^- \dots
\end{equation}
The two branches are coupled by an infinitesimal evolution operation at the final time $t_f$ (see~\cite{Kamenev11}
for more details on the discretized action).


For later convenience, in the following we will split the real-time action into the sum of three types of terms. The free ones, that we will can
$S_o$, the ones that {\it via} the potential $V$ depend on the disorder,  and the ones that depend on the sources
${\bf y}$ that we simply call $S_{\bf y}[{\bf s}^+, {\bf s}^-]$. The weight is, therefore,
\begin{equation}
S_o[{\bf s}^+]-S_o[{\bf s}^{-}] + S_V[{\bf s}]-S_V[{\bf s}^{-}] + S_{\bf y}[{\bf s}^+, {\bf s}^-]
\; .
\end{equation}

We wish to consider the real-time dynamics
of the system, starting from an equilibrium initial
condition at the initial time, $t_0=0$. We then assume that $\hat{\rho}(0)$ is
given by
\begin{equation}
\hat{\rho}(0)=\hat{\rho}_{\rm eq}(0)={\mathcal Z}_{\rm eq}^{-1}(\beta) \ e^{-\beta \hat H_0}
\end{equation}
taking the Gibbs-Boltzmann equilibrium form
at temperature $T\equiv 1/\beta$.
Given that the initial
state is thermalised at temperature $T$, it is correlated with
disorder, that is, $\hat{\rho}_{\rm eq}(0)$ is disorder dependent.
To be more precise, the generating functional without sources
\begin{equation}
{\mathcal Z}\equiv {\mathcal Z}[0,0] = \mbox{Tr} \, \hat{\rho}(0) =1
\end{equation}
is normalized by construction, while the exponential factor $e^{-\beta \hat H_0}$ and the normalisation  constant or partition function
$
{\mathcal Z}_{\rm eq}(\beta)
$
are both disorder dependent.

Finally, representing the
matrix element $\langle {\bf s}^+|e^{-\beta \hat H_0}|{\bf s}^-\rangle$ as
a path-integral in imaginary time, that we call $\tau$, ${\mathcal Z}[{\bf y}^+,{\bf y}^{-}]$ can be written as
\begin{eqnarray}
{\mathcal Z}[{\bf y}^+,{\bf y}^{-}]
\! & \!\! = \!\! & \!
\int D{\bf s}^+\, D{\bf s}^-
\;
e^{{{\rm i}\over \hbar}
\left( S_o[{\bf s}^+]-S_o[{\bf s}^-]-\int dt \; (V[{\bf s}^+(t),J]-
V[{\bf s}^-(t),J])+S_{\bf y}[{\bf s}^+,{\bf s}^-]\right)}
\nonumber\\
&& \qquad\qquad \times \; {1\over {\mathcal Z}_{\rm eq}(\beta)}
\int_{{\bf s}^e(0)={\bf s}^-(0)}^{{\bf s}^e(-\beta\hbar)
={\bf s}^+(0)} D{\bf s}^e
\;
e^{-{1\over \hbar}\left(
S_o[{\bf s}^e]+\int_0^{-\beta\hbar} d\tau\, V[{\bf s}^e(\tau),J_0]
\right)}
\; .
\label{gf2}
\end{eqnarray}
Equation~(\ref{gf2}) clearly shows that, for each realisation of disorder,
the imaginary and real time integration paths are correlated via
the boundary conditions and, therefore, they do not factorize. Performing the
average over the disorder further couples the real and imaginary
time degrees of freedom, leading to mixed two-time correlation functions, as
we will now show.

 \subsection{The average over disorder}

The factor $1/{\mathcal Z}_{\rm eq}(\beta)$ in Eq.~(\ref{gf2}) obliges us to introduce
replicas to perform the quenched average. One possible way
is to use use the relation $1/{\mathcal Z}_{\rm eq}(\beta)=\lim_{n\to 0}({\mathcal Z}_{\rm eq}(\beta))^{n-1}$
and to replicate the system only at $t_0=0$.
Instead, we follow here the same route as in Refs.~\cite{CuLoNe17,CuLoNePiTa18,HoJaYo83,FrPa95,BaBuMe96}:
we start from the average of the logarithm of the full generating functional ${\mathcal Z}[{\bf y}^+, {\bf y}^-]$ and we replicate
the system at all times via the relation
\begin{equation}
[\ln {\mathcal Z}]=\lim_{n\to 0}{1\over n} \left( [{\mathcal Z}^n] -1 \right)
\; .
\end{equation}
The disorder averaged replicated functional $[{\mathcal Z}^n]$ reads
\begin{equation}
[{\mathcal Z}^n]=\int\prod_{a=1}^n D{\bf s}^+_a D{\bf s}^-_a
 D{\bf s}^e_a \;\;  e^{{{\rm i}\over \hbar}S_{\rm eff}} ,
\label{zaidbar}
\end{equation}
where the full effective action $S_{\rm eff}$ is given by
\begin{eqnarray}
S_{\rm eff}
\! & \!\! = \!\! & \!
\sum_a
S_o[{\bf s}^+_a]-S_o[{\bf s}^-_a]+S_{\bf y}[{\bf s}^+_a,{\bf s}^-_a]
+{\rm i} S_o[{\bf s}^e_a]\nonumber\\ & &
+{{\rm i} J^2\over 4\hbar} N\sum_{ab}\int dt\int dt'\left[
\left({1\over N}{\bf s}^+_a(t)\cdot {\bf s}^+_b(t')\right)^p +
\left({1\over N}{\bf s}^-_a(t)\cdot {\bf s}^-_b(t')\right)^p
-2\left({1\over N}{\bf s}^+_a(t)\cdot {\bf s}^-_b(t')\right)^p\right]
\nonumber\\& &
+{J_0 J\over 2\hbar}  N\sum_{ab}\int dt\int d\tau \left[
\left({1\over N}{\bf s}^+_a(t)\cdot {\bf s}^e_b(\tau)\right)^p
-\left({1\over N}{\bf s}^-_a(t)\cdot {\bf s}^e_b(\tau)\right)^p
\right]\nonumber\\& &
-{{\rm i}J_0^2\over 4\hbar}  N\sum_{ab}\int d\tau \int d\tau_1
\left({1\over N}{\bf s}^e_a(\tau)\cdot {\bf s}^e_b(\tau_1)\right)^p .
\label{seff}
\end{eqnarray}
The real time integrals in $S_{\rm eff}$ run from the initial time $t=0$ to the final one $t_f$. The 
imaginary time integral runs from $0$ to $-\beta \hbar$ and all functions of $\tau$ are periodic, 
with period $\beta\hbar$.
The averaging over disorder not only induced interactions between different replicas, as
usually found in the evaluation of averaged replicated partition functions, but it also
generated interactions between variables living on different time branches. Indeed,
the interactions between variables living on the two real time axes, variables living on the
real-time axes and the imaginary time one, and variables defined on the imaginary time
axis are explicit in this effective action.

\subsection{Introduction of auxiliary functions}

In order to write the effective action~(\ref{seff}) in a
more compact form, we define the operator
\begin{equation}
{\bf Op}_{ab}(\hat{t},\hat{t}')= \left(
\begin{array}{ccc}
{\rm Op}^{++}_{ab}(t,t')& {\rm Op}^{+-}_{ab}(t,t')& 0 \\
{\rm Op}^{-+}_{ab}(t,t')& {\rm Op}^{--}_{ab}(t,t')& 0 \\
0 & 0& {\rm Op}^{ee}_{ab}(\tau,\tau')
\end{array}
\right)=\left( {\rm Op}_{ab}^{\alpha\beta}(\hat{t},\hat{t}') \right) ,
\end{equation}
where
$(a,b)$
and $(\alpha,\beta)$ are replica, $a,b=1, \dots, n$, and time-branch $\alpha, \beta=+,-,e$ indices,
respectively. The entries of the above matrix are then given by
\begin{eqnarray}
{\rm Op}^{++}_{ab}(t,t')&=& \delta_{ab}\left[\left(m{\partial_t^2}+
z^+_a(t)\right)\delta(t-t')\right]  \; ,
\nonumber\\
{\rm Op}^{+-}_{ab}(t,t')&=& 0 \; ,
\nonumber\\
{\rm Op}^{-+}_{ab}(t,t')& =& 0  \; ,
\nonumber\\
{\rm Op}^{--}_{ab}(t,t')& =& -\delta_{ab} \left(m{\partial_t^2}+
z^-_a(t)\right)\delta(t-t')  \; ,
\nonumber\\
{\rm Op}_{ab}^{ee}(\tau,\tau')&=& {\rm i} \delta_{ab}\left(m{\partial_{\tau}^2}-z^e_a(\tau)\right)\delta(\tau-\tau')
\; ,
\end{eqnarray}
and they can be symmetrized with respect to the time indices after simple
integrations by parts.
We now introduce the identity
\begin{eqnarray}
1&=& \int\prod_{\alpha\beta}\prod_{ab} D Q_{ab}^{\alpha\beta}
\,\,\delta\left( {1\over N}{\bf s}^{\alpha}_a(\hat{t})\cdot
{\bf s}_b^{\beta}(\hat{t}')-Q_{ab}^{\alpha\beta}(\hat{t},\hat{t}')
\right)
\nonumber\\
&\propto & \int\prod_{\alpha\beta}\prod_{ab} D Q_{ab}^{\alpha\beta}
\, D\lambda_{ab}^{\alpha\beta}\exp\left[
{-{\rm i}\over 2\hbar}\sum_{ab}\int d\hat{t}\, d\hat{t}'\left(
{\bf s}^{\alpha}_a(\hat{t})\cdot{\bf s}_b^{\beta}(\hat{t}')
 -N Q_{ab}^{\alpha\beta}(\hat{t},\hat{t}')\right) \lambda_{\alpha\beta}(\hat t, \hat t') \right]
\end{eqnarray}
where the auxiliary matrix elements $\lambda_{ab}(t,t')$ should satisfy symmetry conditions,
and we insert it in Eq.~(\ref{zaidbar}) to get the following
expression for the effective action:
\begin{eqnarray}
S_{\rm eff} &= & -{1\over 2}\sum_{ab}\int d\hat{t}\, d\hat{t}'
\,{\bf s}_a^{\alpha}(\hat{t})\cdot \left(
{\rm Op}_{ab}^{\alpha\beta}(\hat{t},\hat{t}')+\lambda_{ab}^{\alpha\beta}
(\hat{t},\hat{t}')\right)\cdot {\bf s}_b^{\beta}(\hat{t}')\nonumber\\
& & +{N\over 2}\sum_{ab}\int d\hat{t}\, d\hat{t}'
\,\lambda_{ab}^{\alpha\beta}(\hat{t},\hat{t}')
\, Q_{ab}^{\alpha\beta}(\hat{t},\hat{t}')\nonumber\\& &
+{N\over 2}\sum_a\int dt \left(z^+_a(t)-z^-_a(t)\right)
-{\rm i}{N\over 2}\sum_a\int d\tau\, z^e_a(\tau)\nonumber\\& &
+{{\rm i}\over 4\hbar}{J}^2 N \sum_{ab}\int dt\, dt' \left[
\left(Q^{++}_{ab}(t,t')\right)^p+\left(Q^{--}_{ab}(t,t')\right)^p-
\left(Q^{+-}_{ab}(t,t')\right)^p-\left(Q^{-+}_{ab}(t,t')\right)^p
\right]\nonumber\\ & &
+{1\over 4\hbar}J_0 J N \sum_{ab}\int dt\, d\tau \left[
\left(Q^{+e}_{ab}(t,\tau)\right)^p+\left(Q^{e+}_{ab}(\tau,t)\right)^p-
\left(Q^{-e}_{ab}(t,\tau)\right)^p-\left(Q^{e-}_{ab}(\tau,t)\right)^p
\right]\nonumber\\ & &
-{{\rm i}\over 4\hbar}{J^2_0} N \sum_{ab}\int d\tau\, d\tau_1
\left(Q_{ab}^{ee}(\tau,\tau_1)\right)^p ,
\label{seff1}
\end{eqnarray}
where, when appear repeated,  the sum convention over $\alpha, \beta=+,-,e$ is
assumed. Note that the above expressions are written in a
very condensed form useful to derive the dynamical equations.
Some remarks are needed at this stage in order to
clarify the notation used. The times $\hat{t}$ and $\hat{t}'$
can either stand for real or imaginary time, depending on the
index $\alpha$ or $\beta$ to which they correspond. More
precisely, for $\alpha, \beta = +,-$, the corresponding time
is real, whereas $\alpha, \beta = e$ corresponds to imaginary time.
Let us give an example to illustrate the above
statements: for $\alpha=+$ and $\beta=e$, the first term in
Eq.~(\ref{seff1}) is, by definition, identical to
\begin{eqnarray}
& &-{1\over 2}\sum_{ab}\int d\hat{t}\, d\hat{t}'
\,{\bf s}_a^{+}(\hat{t})\cdot \left(
{\rm Op}_{ab}^{+e}(\hat{t},\hat{t}')+\lambda_{ab}^{+e}
(\hat{t},\hat{t}')\right) {\bf s}_b^{e}(\hat{t}')\nonumber\\
& &\qquad\qquad
\equiv
-{1\over 2}\sum_{ab}\int dt\, d\tau
\,{\bf s}_a^{+}(t)\cdot \left(
{\rm Op}_{ab}^{+e}(t,\tau)+\lambda_{ab}^{+e}
(t,\tau)\right) {\bf s}_b^{e}(\tau)
\; .
\end{eqnarray}
Moreover, the operator ${\rm Op}$ and the auxiliary function $\lambda$
are naturally symmetrised though we do not write this operation explicitly
so as to lighten the notation.

\subsection{Saddle-point and expected values}

At the saddle point, the expected values of the order parameters,
$Q_{ab}^{\alpha\beta}(\hat{t},\hat{t}')$ are given by
\begin{equation}
NQ^{++}_{ab}(t,t')=[{\langle{\bf s}^+_a(t)\cdot {\bf s}^+_b(t')\rangle}]
=N\left(C_{ab}(t,t')-{{\rm i}\hbar\over 2}(R_{ab}(t,t')+R_{ab}(t',t))\right) ,
\label{op1}
\end{equation}
\begin{equation}
NQ^{+-}_{ab}(t,t')=[{\langle{\bf s}^+_a(t)\cdot {\bf s}^-_b(t')\rangle}]
=N\left(C_{ab}(t,t')+{{\rm i}\hbar\over 2}(R_{ab}(t,t')-R_{ab}(t',t))\right)  ,
\label{op2}
\end{equation}
\begin{equation}
NQ^{-+}_{ab}(t,t')=[{\langle{\bf s}^-_a(t)\cdot {\bf s}^+_b(t')\rangle}]
=N\left(C_{ab}(t,t')-{{\rm i}\hbar\over 2}(R_{ab}(t,t')-R_{ab}(t',t))\right)  ,
\label{op3}
\end{equation}
\begin{equation}
NQ^{--}_{ab}(t,t')=[{\langle{\bf s}^-_a(t)\cdot {\bf s}^-_b(t')\rangle}]
=N\left(C_{ab}(t,t')+{{\rm i}\hbar\over 2}(R_{ab}(t,t')+R_{ab}(t',t))\right)  ,
\label{op4}
\end{equation}
for the ones with support on the real time axis;
\begin{equation}
NQ^{+e}_{ab}(t,\tau)=
[{\langle{\bf s}^+_a(t)\cdot {\bf s}^e_b(\tau)\rangle}]
\; ,
\label{op5}
\end{equation}
\begin{equation}
NQ^{-e}_{ab}(t,\tau)=
[{\langle{\bf s}^-_a(t)\cdot {\bf s}^e_b(\tau)\rangle}]
\; ,
\label{op6}
\end{equation}
\begin{equation}
NQ^{e+}_{ba}(\tau,t) = [{\langle{\bf s}^e_b(\tau)\cdot {\bf s}^+_a(t)\rangle}]
\; ,
\label{op7}
\end{equation}
\begin{equation}
NQ^{e-}_{ba}(\tau,t) = [{\langle{\bf s}^e_b(\tau)\cdot {\bf s}^-_a(t)\rangle}]
\; ,
\label{op8}
\end{equation}
for the ones with support on the real time and imaginary time axes;
and
\begin{equation}
NQ^{ee}_{ab}(\tau,\tau')=
[\langle{\bf s}^e_a(\tau)\cdot {\bf s}^e_b(\tau')\rangle]
\; ,
\label{op7}
\end{equation}
for the ones with support on the imaginary time axis.
In the above expressions, $C_{ab}(t,t')$ and $R_{ab}(t,t')$ are defined as
\begin{eqnarray}
&&
C_{ab}(t,t')\equiv {1\over 2N} \; [\langle{\bf s}^+_a(t)\cdot
{\bf s}^-_b(t')+{\bf s}^-_a(t)\cdot {\bf s}^+_b(t')\rangle]
= \frac{1}{2} [ Q_{ab}^{+-}(t,t') + Q_{ab}^{-+}(t,t')]
\; ,
\nonumber\\
&&
R_{ab}(t,t')\equiv {{\rm i}\over \hbar N} \; [\langle{\bf s}^+_a(t)\cdot
\left({\bf s}^+_b(t')-{\bf s}^-_b(t')\right)\rangle]
=
\frac{{\rm i}}{\hbar} [ Q_{ab}^{++}(t,t') + Q_{ab}^{+-}(t,t')]  \;  ,
\end{eqnarray}
and one can easily check that the functions
$Q^{(\pm)(\pm)}_{ab}(t,t')$ satisfy the identity
\begin{equation}
Q^{++}_{ab}(t,t')+
Q^{--}_{ab}(t,t')-Q^{+-}_{ab}(t,t')-Q^{-+}_{ab}(t,t')=0
\; .
\label{forq0}
\end{equation}

The functional integration over the variables ${\bf s}^+_a(t)$,
${\bf s}^-_a(t)$ and ${\bf s}^e_a(\tau)$ is now quadratic and can be
performed. This amounts to replacing the quadratic term in
$i/\hbar \; S_{\rm eff}$ by
\begin{equation}
-{N\over 2} {\mbox{Tr}}\ln\left( {{\rm i}\over \hbar}{\bf Op}_{ab}
(\hat{t},\hat{t}')+{{\rm i}\over \hbar} {\bf L}_{ab}
(\hat{t},\hat{t}')\right) ,
\end{equation}
where the matrix ${\bf L}_{ab}(\hat{t},\hat{t}')$ is
defined as the symmetrized version of
\begin{equation}
{\bf L}_{ab}(\hat{t},\hat{t}')=\left(
\begin{array}{ccc}
\lambda_{ab}^{++}(t,t')&\lambda_{ab}^{+-}(t,t')&\lambda_{ab}^{+e}(t,\tau')
\\
\lambda_{ab}^{-+}(t,t')&\lambda_{ab}^{--}(t,t')&\lambda_{ab}^{-e}(t,\tau')
\\
\lambda_{ab}^{e+}(\tau,t')&\lambda_{ab}^{e-}(\tau,t')&
\lambda_{ab}^{ee}(\tau,\tau')
\end{array}
\right)
\end{equation}
and the differential operator has to be conveniently symmetrized as well.
The effective action is now proportional to $N$ and, in the
large $N$ limit, this allows us to evaluate the
functional~(\ref{zaidbar}) by the steepest descendent method. It is
easier to write the equations in matricial form. With this aim in mind, we encode $Q_{ab}^{\alpha\beta}(\hat{t},\hat{t}')$
in the matrix
\begin{equation}
{\bf Q}_{ab}(\hat{t},\hat{t}') \equiv \left(
\begin{array}{ccc}
Q_{ab}^{++}(t,t')&Q_{ab}^{+-}(t,t')&Q_{ab}^{+e}(t,\tau')
\\
Q_{ab}^{-+}(t,t')&Q_{ab}^{--}(t,t')&Q_{ab}^{-e}(t,\tau')
\\
Q_{ab}^{e+}(\tau,t')&Q_{ab}^{e-}(\tau,t')&
Q_{ab}^{ee}(\tau,\tau')
\end{array}
\right)  ,
\end{equation}
and we define
\begin{equation}
F[{\bf Q}]_{ab}(\hat{t},\hat{t}')
\equiv
\frac{p}{2\hbar}
\left(
\begin{array}{ccc}
+ J^2 \left(Q_{ab}^{++}(t,t')\right)^{p-1}&
- J^2 \left(Q_{ab}^{+-}(t,t')\right)^{p-1}&
- {\rm i} J J_0 \left(Q_{ab}^{+e}(t,\tau')\right)^{p-1}\\
- J^2 \left(Q_{ab}^{-+}(t,t')\right)^{p-1}&
+ J^2 \left(Q_{ab}^{--}(t,t')\right)^{p-1}&
+ {\rm i} J J_0 \left(Q_{ab}^{-e}(t,\tau')\right)^{p-1}\\
- {\rm i} J J_0 \left(Q_{ab}^{e+}(\tau,t')\right)^{p-1}&
+ {\rm i} J J_0 \left(Q_{ab}^{e-}(\tau,t')\right)^{p-1}&
-J_0^2 \left(Q_{ab}^{ee}(\tau,\tau')\right)^{p-1}
\end{array}
\right)  \; .
\end{equation}
The saddle-point equations with respect to $\lambda_{ab}^{\alpha\beta}
(\hat{t},\hat{t}')$ and $Q_{ab}^{\alpha\beta}(\hat{t},\hat{t}')$
yield
\begin{equation}
{\bf L}_{ab}(\hat{t},\hat{t}')=-{\bf Op}_{ab}(\hat{t},\hat{t}')+
{\hbar\over {\rm i}}\left({\bf Q}^{-1}\right)_{ab}(\hat{t},\hat{t}')
\label{spl}
\end{equation}
and
\begin{equation}
{\bf L}_{ab}(\hat{t},\hat{t}')=-{{\rm i}}
F[{\bf Q}]_{ab}(\hat{t},\hat{t}') \;  ,
\label{spq}
\end{equation}
respectively. Equations~(\ref{spl}) and~(\ref{spq}) imply
\begin{eqnarray}
{{\rm i}\over \hbar}\sum_c\int d\hat{t}_1\, {\bf Op}_{ac}
(\hat{t},\hat{t}_1) {\bf Q}_{cb}(\hat{t}_1,\hat{t}') &=&
{\bf I} \ \delta_{ab}\delta(\hat{t}-\hat{t}')
-{1\over \hbar}\sum_c\int\, d\hat{t}_1 \
F[{\bf Q}]_{ac}(\hat{t},\hat{t}_1){\bf Q}_{cb}(\hat{t}_1,\hat{t}')
\; ,
\label{fspe}
\\
{{\rm i}\over \hbar}\sum_c\int d\hat{t}_1 \
{\bf Q}_{ac}(\hat{t},\hat{t}_1)
{\bf Op}_{cb}(\hat{t}_1,\hat{t}') &=&
{\bf I} \
\delta_{ab}\delta(\hat{t}-\hat{t}')
-{1\over \hbar}\sum_c\int\, d\hat{t}_1 \
{\bf Q}_{ac}(\hat{t},\hat{t}_1)
F[{\bf Q}]_{cb}(\hat{t}_1,\hat{t}')
\; ,
\label{fspe-left}
\end{eqnarray}
after multiplying, operationally, by ${\rm i}{\bf Q}/\hbar$ on the right and on the left, respectively.
The symbol ${\bf I}$ represents the identity in the $\alpha,\beta$ indices.
Finally, the saddle point
equations with respect to $z^+_a(t)$, $z^-_a(t)$ and $z^e_a(\tau)$
give back, as expected, the spherical constraints,
\begin{equation}
Q_{aa}^{++}(t,t)=Q_{aa}^{--}(t,t)=Q_{aa}^{ee}(\tau,\tau)=1
\; .
\end{equation}

\subsection{The dynamic equations}

A set of dynamic equations, in which the time derivatives act on the first time argument,
follow from Eq.~(\ref{fspe}) and the definitions of the order parameters given by Eqs.~(\ref{op1})-(\ref{op7}).
Moreover, since the response is causal, that is,
$R_{ab}(t,t')=0$ for $t\le t'$, one is naturally lead to the
identity
\begin{equation}
{1\over N}\overline{{\partial \langle{\bf s}^e_a(\tau)\rangle\over
\partial {\bf f}_b(t)}} \vert_{{\bf f}=0} = 0
\; ,
\end{equation}
where ${\bf f}(t)$ is a small perturbation applied at
time $t$. The above equation implies
\begin{equation}
Q_{ab}^{e+}(\tau,t)
=
Q_{ab}^{e-}(\tau,t)
\; .
\end{equation}
The equations of motion for the response function, $R_{ab}(t,t')$, and
auto-correlation function, $C_{ab}(t,t')$ follow
from the subtraction of the $++$ and $+-$, and the $+-$ and $-+$
components of Eq.~(\ref{fspe}), respectively; they read
\begin{eqnarray}
& &\left(m{\partial_t^2} + z^+_a(t)\right)
R_{ab}(t,t')=\delta_{ab}
\delta(t-t')
-{J^2p\over 2{\rm i}\hbar}\sum_c\int dt_1\, \left[
\left(Q^{++}_{ac}(t,t_1)\right)^{p-1}-
\left(Q^{+-}_{ac}(t,t_1)\right)^{p-1}\right] R_{cb}(t_1,t')
\nonumber
\label{rtt1}
\end{eqnarray}
and
\begin{eqnarray}
& &\left(m{\partial_t^2}+{z^+_a(t)+z^-_a(t)\over 2}
\right)C_{ab}(t,t')+{{\rm i}\hbar\over 4}\left(z^+_a(t)-z^-_a(t)\right)
(R_{ab}(t,t')-R_{ab}(t,t'))\nonumber\\& &
\nonumber\\& &
\qquad
=-{J^2p\over 2\hbar}\sum_c\int dt_1\, {\rm Im}
\left[\left(Q^{++}_{ac}(t,t_1)\right)^{p-1}Q_{cb}^{+-}(t_1,t')
-\left(Q^{+-}_{ac}(t,t_1)\right)^{p-1}Q_{cb}^{--}(t_1,t')\right]\nonumber\\
& &
\qquad\qquad +{J_0Jp\over 2\hbar}\sum_c\int d\tau\, \left({G^\lceil_{ac}}
(t,\tau)\right)^{p-1}{G^\rceil_{cb}}(\tau,t')
\; .
\label{ctt1}
\end{eqnarray}
The imaginary part of eq.~(\ref{ctt1}) implies
\begin{equation}
z_a(t)\equiv z^+_a(t)= z^-_a(t)
\; .
\label{zequacao}
\end{equation}
At this point, it is convenient to introduce the
following identities (see ref.~\cite{CuLo99} for details
on their derivation)
\begin{equation}
\left(Q^{++}_{ab}(t,t')\right)^{p-1}-
\left(Q^{+-}_{ab}(t,t')\right)^{p-1}=
2i{\rm Im}\left[
C_{ab}(t,t')-{{\rm i}\hbar\over 2}R_{ab}(t,t')\right]^{p-1}
\; ,
\label{ident1}
\end{equation}
\begin{equation}
\left(Q^{++}_{ab}(t,t')\right)^{p-1}-
\left(Q^{-+}_{ab}(t,t')\right)^{p-1}=
2i{\rm Im}\left[
C_{ab}(t,t')-{{\rm i}\hbar\over 2}R_{ab}(t',t)\right]^{p-1}
\; ,
\label{ident11}
\end{equation}
\begin{equation}
\left(Q^{-+}_{ab}(t,t')\right)^{p-1}-
\left(Q^{--}_{ab}(t,t')\right)^{p-1}=
2i{\rm Im}\left[
C_{ab}(t,t')-{{\rm i}\hbar\over 2}R_{ab}(t,t')\right]^{p-1}
\; ,
\label{ident2}
\end{equation}
\begin{eqnarray}
& &{\rm Im}\left[\left(Q^{++}_{ac}(t,t_1)\right)^{p-1}Q_{cb}^{+-}(t_1,t')
-\left(Q^{+-}_{ac}(t,t_1)\right)^{p-1}Q_{cb}^{--}(t_1,t')\right]=
\nonumber\\& &
\qquad\qquad\qquad
2C_{cb}(t_1,t'){\rm Im}\left[
C_{ac}(t,t_1)-{{\rm i}\hbar\over 2}R_{ac}(t,t_1)\right]^{p-1}
\nonumber\\& &
\qquad\qquad\qquad
-\hbar R_{cb}(t',t_1){\rm Re}
\left[
C_{ac}(t,t_1)-{{\rm i}\hbar\over 2}(R_{ac}(t,t_1)+R_{ac}(t_1,t))\right]^{p-1}
\label{ident3}
\end{eqnarray}
and also
\begin{equation}
Q_{ab}^{++}(t,t')-Q_{ab}^{-+}(t,t')=-{\rm i}\hbar R_{ab}(t',t)
\; ,
\label{ident4}
\end{equation}
\begin{equation}
Q_{ab}^{+-}(t,t')-Q_{ab}^{--}(t,t')=-{\rm i}\hbar R_{ab}(t',t)
\; ,
\label{ident5}
\end{equation}
which can be easily obtained from the definitions of the
real-time order parameters given by Eqs.~(\ref{op1})-(\ref{op4}).
Using Eqs.~(\ref{zequacao}), (\ref{ident1}) and (\ref{ident3}),
the equations for $R_{ab}(t,t')$ and $C_{ab}(t,t')$ can be written
as
\begin{eqnarray}
& &\left(m{\partial_t^2} + z_a(t)\right)
R_{ab}(t,t')=\delta_{ab}
\delta(t-t')
-{J^2p\over \hbar}\sum_c\int dt_1\,
R_{cb}(t_1,t')\,{\rm Im} \left[C_{ac}(t,t_1)-{{\rm i}\hbar\over 2}
R_{ac}(t,t_1)\right]^{p-1}
\label{rtt2}
\end{eqnarray}
and
\begin{eqnarray}
& &\left(m{\partial^2_t}
+z_a(t)\right)C_{ab}(t,t')
=-{J^2p\over 2\hbar}\sum_c\int dt_1\, \left\{
2C_{cb}(t_1,t'){\rm Im}\left[
C_{ac}(t,t_1)-{{\rm i}\hbar\over 2}R_{ac}(t,t_1)\right]^{p-1}\right.
\nonumber\\
& &
\qquad\qquad\qquad\qquad\qquad\qquad
\left.-\hbar R_{cb}(t',t_1){\rm Re}
\left[
C_{ac}(t,t_1)-{{\rm i}\hbar\over 2}(R_{ac}(t,t_1)+R_{ac}(t_1,t))\right]^{p-1}
\right\}
\nonumber\\& &
\qquad\qquad\qquad\qquad\qquad\qquad
+{J_0Jp\over 2\hbar}\sum_c\int d\tau\,
\left({G^\lceil_{ac}}(t,\tau)\right)^{p-1}
{G^\rceil_{cb}}(\tau,t')
\; ,
\label{ctt2}
\end{eqnarray}
respectively.

We now discuss the computation of the evolution equations
for ${G^\lceil_{ab}}(t,\tau)$, ${G^\rceil_{ab}}(\tau,t)$ and $Q^{ee}_{ab}(\tau,\tau')$.
From here on, we will use the notation
\begin{equation}
Q_{ab}(\tau,\tau')\equiv Q^{ee}_{ab}(\tau,\tau')
\; .
\end{equation}
Using Eqs.~(\ref{zequacao}), (\ref{ident1}) and (\ref{ident2}),
the equations for the $+e$ and $-e$ components of Eq.~(\ref{fspe})
collapse, as they should, into a single equation which gives
the evolution in real time of ${G^\lceil_{ab}}(t,\tau)$; it
reads
\begin{eqnarray}
& &
\left(m{\partial^2_t}+z_a(t)\right){G^\lceil_{ab}}(t,\tau)
=-{J_0 J p\over \hbar}\sum_c\int dt_1\,
 {\rm Im}\left[
C_{ac}(t,t_1)-{{\rm i}\hbar\over 2}R_{ac}(t,t_1)\right]^{p-1}
\
{G^\lceil_{cb}}(t_1,\tau)
\nonumber\\
& &
\qquad\qquad\qquad\qquad\qquad\;\;\;\;\;\;
+{J_0^2 p\over 2\hbar}\sum_c\int d\tau_1\,
\left({G^\lceil_{ac}}(t,\tau_1)\right)^{p-1}
Q_{cb}(\tau_1,\tau)
\; .
\label{rqbar}
\end{eqnarray}
Moreover, utilising the identities~(\ref{ident4}) and (\ref{ident5}),
the equations for the $e+$ and $e-$ components of Eq.~(\ref{fspe})
also collapse into a single equation which describes now the
evolution in imaginary time of $G^\rceil_{ab}(\tau,t)$; 
since it will not be very useful to implement a numerical code to 
construct the solution step-by-step in real time, we do not write
it explicitly here.

Another set of dynamic equations, derived from Eq.~(\ref{fspe-left}), where the time-derivatives act on the second time
argument in the correlation functions and linear responses can also be derived.  Usually,
these equations are just compatible with the ones already derived above and
do not yield additional information nor are particularly useful. In this case, however, the numerical solution of the
problem will benefit from an important simplification if one of these equations is used. This is the equation in which
the real time differential operator acts on ${G^\rceil_{cb}}(\tau,t)$, see Sec.~\ref{sec:conclusions}.
For this reason, we write this equation explicitly,
\begin{eqnarray}
{\rm i} \left(m{\partial_t^2}+z_a(t)\right) G^\rceil_{ab}(\tau,t) &=&
- J^2\sum_c \int dt_1 \ G^\rceil_{ac}(\tau,t_1) \left[ \left( Q_{cb}^{++} (t_1,t) \right)^{p-1}- \left( Q_{cb}^{-+}(t_1,t) \right)^{p-1} \right]
\nonumber\\
&&
+{\rm i} 
\frac{J_0Jp}{2\hbar}  \sum_c \int d\tau_1 \; Q_{ac}(\tau,\tau_1) \left( G^\rceil_{ac}(\tau_1,t) \right)^{p-1}
\; ,
\end{eqnarray}
that becomes
\begin{eqnarray}
\left(m{\partial_t^2}+z_a(t)\right) G^\rceil_{ab}(\tau,t)
&= &
- \frac{J^2}{\hbar} 
\sum_c \int dt_1 \ G^\rceil_{ac}(\tau,t_1) \; 2 \mbox{Im} \left[ C(t,t_1) - \frac{{\rm i}\hbar}{2} R(t,t_1) \right]^{p-1}
\nonumber\\
&&
+ \frac{J_0Jp}{2\hbar} \sum_c \int d\tau_1 \; Q_{ac}(\tau,\tau_1) \left( G^\rceil_{ac}(\tau_1,t) \right)^{p-1}
\; .
\end{eqnarray}
In the last step we used the identity (\ref{ident11}), that reverses the time ordering of the response in the integrand and
ensures in this way that the equation is causal in real time.

Finally, the $ee$ component of Eq.~(\ref{fspe}) yields the equation
of motion for $Q_{ab}(\tau,\tau')$, which reads
\begin{equation}
\left(m{\partial_\tau^2}-z^e_a(\tau)\right)
Q_{ab}(\tau,\tau')=-\delta(\tau-\tau')
-{J_0^2p\over 2\hbar}\sum_c\int d\tau_1\,
\left(Q_{ac}(\tau,\tau_1)\right)^{p-1}  Q_{cb}(\tau_1,\tau')
\; .
\label{staticq}
\end{equation}
As can be seen, Eq.~(\ref{staticq}) decouples from
all the other equations and yields the static properties of the
model. Since it describes the system in equilibrium,
$Q_{ab}(\tau,\tau')$ and $z^e_a(\tau)$ can be taken to be imaginary-time-translation
invariant and imaginary-time independent, respectively; assuming further that
$z^e_a$ does not depend on the replica index
Eq.~(\ref{staticq}) can be written as
\begin{equation}
\left(m{\partial_\tau^2}-z^e\right)
Q_{ab}(\tau)=-\delta(\tau)
-{J_0^2p\over 2\hbar}\sum_c\int d\tau_1\;
\left(Q_{ac}(\tau-\tau_1)\right)^{p-1}  Q_{cb}(\tau_1)
\; ,
\label{staticq1}
\end{equation}
which is equivalent to the equilibrium equation derived in Refs.~\cite{CuGrSa00,CuGrSa01}, with the 
same sign convention.
Moreover, the classical limit of the above full set of equations yields the equations in~\cite{CuLoNe17} as we will 
show explicitly below.

In none of these equations we wrote the limits of the real and imaginary time integrals. These are simple 
to guess and we will make these limits explicit
in the following sections. One can, however, already check at this stage that all equations respect causality.

 \subsection{The replica structure}

 On the one hand, the replica index structure of all the two-time functions will be dictated by the one chosen by the
 initial conditions. It is easy to see that, indeed, the replica structure gets propagated by the dynamic equations.
 On the other hand, the replica structure of the ${Q^{ee}_{ab}}(\tau-\tau')$ imaginary time dependent matrix
will be different depending on the initial conditions chosen.

 In the paramagnetic phase, the imaginary-time dependent replica matrix order parameter is symmetric and
 diagonal meaning that
 \begin{equation}
 {Q^{\rm rs}_{ab}}(\tau,\tau') = \delta_{ab} \ q_d(\tau-\tau')
 \; .
 \end{equation}
 In the ordered phase relevant in equilibrium below the curve $(T_s, \Gamma_s)$,
 with $\Gamma=\hbar^2/(Jm)$,
 the elements of the replica matrix order parameter are imaginary time independent apart from
 the ones on the diagonal. Moreover, the matrix has a
 one-step level of replica symmetry breaking:
 \begin{equation}
 Q^{\rm 1rsb}_{ab}(\tau,\tau') = \delta_{ab} \ [q_d(\tau-\tau') -q] + \epsilon_{ab} \ q
 \; ,
 \end{equation}
 with $\epsilon_{ab}=1$ for $a,b$ within diagonal blocks of size $m$ and  $\epsilon_{ab}=0$ otherwise.
The equations that determine the parameters $q,m,q_d(\tau)$ can be found in~\cite{CuGrSa00,CuGrSa01} and we do not reproduce them here.

As it is well known, systems with one step replica symmetry breaking structure in the ordered phase
have a narrow region of temperatures in the classical limit, and temperature and quantum control 
parameter, say $\Gamma$, in the quantum one, where a large number of metastable states combine in the 
replica solution into a replica symmetric structure. This occurs in sector of the phase diagram $(T,\Gamma)$ 
which goes beyond $(T_s, \Gamma_s)$ and until a {\it dynamical critical line} $(T_d, \Gamma_d)$
in the quantum problem~\cite{CuGrSa00,CuGrSa01,BiCu01,Bapst}. Quenches from equilibrium in this 
region are particularly interesting since they keep a simple structure of initial conditions while have the 
possibility of remaining blocked out of equilibrium after the quench, see Ref.~\cite{CuLoNe17} for the 
classical realisation of this phenomenon. For this reason, in the rest of this paper we focus on the 
dynamic equations for replica symmetric initial conditions, the extension to the one step replica symmetry broken being straightforward.

 \subsection{The dynamic equations for replica symmetric initial conditions}


 After having simplified the replica structure, we call
 \begin{eqnarray}
 C(t,t') &\equiv& (2N)^{-1} \sum_{i=1} \langle s^+_i(t) s^-_i(t') + s^+_i(t') s^-_i(t)\rangle
 \; , \\
 G^\lceil(t,\tau) &\equiv& N^{-1} \sum_{i=1} \langle s_i(t) s^e_i(\tau)\rangle
 \; , \\
 {G^\rceil}(\tau,t) &\equiv& N^{-1} \sum_{i=1} \langle s^e_i(\tau) s_i(t)\rangle
 \; , \\
 q_d(\tau,\tau')  &\equiv& N^{-1} \sum_{i=1} \langle s^e_i(\tau) s^e_i(\tau')\rangle
 \; ,
 \end{eqnarray}
 and the equations read
 \begin{eqnarray}
 \left( m \partial^2_{t} +z(t) \right) R(t,t') \! & \! = \! & \! \delta(t-t') + \int_{t'}^{t} \! dt_1  \; \Sigma(t,t_1) \, R(t_1,t')
 \; ,
 \label{eq:R}
 \\
\left( m \partial^2_{t} +z(t) \right) C(t,t') \! & \! = \! & \!   \int_0^{t} \! dt_1 \; \Sigma(t,t_1) \, C(t_1,t') + \int_0^{t'} \! dt_1 \; D(t,t_1) \, R(t',t_1)
\nonumber\\
&&
+  \int_0^{\beta\hbar} d\tau_1 \; \Sigma^\lceil(t,\tau_1)  \, G^\rceil (\tau_1,t')
\; ,
\;\;\;\;\;\;
\label{eq:C}
\\
\left( m \partial^2_{t} +z(t) \right) G^\lceil(t,\tau)  \! & \! = \! & \!  \int_0^{t} \! dt_1 \; \Sigma(t,t_1) \, G^\lceil(t_1, \tau)
 +   \int_0^{\beta\hbar} d\tau_1 \; \Sigma^\lceil(t,\tau_1) \, q_d(\tau_1-\tau)
 \; ,
 \label{eq:Qt1tau}
 \\
 \left( m \partial^2_{t} +z(t) \right) {G^\rceil}(\tau, t)
\! & \! = \! & \!
- \int_0^{t} dt_1 \; G^\rceil(\tau,t_1) \, \Sigma(t,t_1) + \int_0^{\beta_0\hbar} d\tau_1 \; q_d(\tau-\tau_1) \, \Sigma^\rceil(\tau_1,t)
 \label{eq:Qtaut1bis}
 \; ,
 \\
\left( m \partial^2_{\tau} - z_e \right) q_d(\tau)  \! & \! = \! & \!  -\delta(\tau)
+  \int_0^{\beta_0\hbar} d\tau_1 \; \Sigma_e(\tau-\tau_1) \, q_d(\tau_1)
\; .
\label{eq:qd}
 \end{eqnarray}
The equation for the Lagrange multiplier (found from $C(t,t)=1$) reads
 \begin{eqnarray}
 z(t) &=&
 \int_0^{t} dt_1 \ [\Sigma(t,t_1) C(t,t_1) + D(t,t_1) R(t,t_1) ] - \left. m \partial_{t}^2 C(t,t') \right|_{t'\to t^-}
 \nonumber\\
 &&
 +  \int_0^{\beta_0\hbar} d\tau_1 \; \Sigma^\lceil(t,\tau_1) \, {G^\rceil}(\tau_1,t)
 \; . 
 \label{eq:z}
 \end{eqnarray}
 The symmetry properties and boundary conditions are
 \begin{eqnarray}
 C(t,t') &=& C(t',t)
 \; ,  \qquad\qquad \qquad\qquad\quad\quad\;\;\;
 C(t,t) = 1
 \; ,  \\
 C(t,0) &=& {G^\lceil}(t,0)
\; ,  \qquad\qquad \qquad\qquad\quad\quad\;\;
 {G^\rceil}(\tau, 0) = q_d(\tau)
 \; ,  \\
 {G^\lceil}(0,\tau) &=& q_d(-\tau) \; = \; q_d(\beta\hbar-\tau)
 \; ,  \qquad\qquad\, 
 R(t, t) = 0 
 \; .
 \end{eqnarray}
 In discrete real time formalism, with $\delta$ the infinitesimal time step,
 the ``next-to-diagonal'' values, or time-derivatives evaluated at
 equal times, are
 \begin{equation}
 C(t+\delta , t) = C(t, t) =1
 \; ,
 \quad
 G^{\lceil}(t+\delta , \tau) = G^{\lceil}(t, \tau)
 \; ,
 \quad
  G^{\rceil}(\tau,t+\delta) = G^{\lceil}(\tau,t)
 \; ,
 \quad\,
  R(t+\delta , t) = 1/m \; .
 \end{equation}
 Finally, the kernels in the integro-differential equations are given by
 \begin{eqnarray}
\Sigma(t,t') &=& - \frac{J^2 p}{\hbar} \ \mbox{Im} \left( C(t,t') - \frac{{\rm i}\hbar}{2} R(t,t') \right)^{p-1}
\!\! ,
\\
D(t,t') &=& \frac{J^2 p}{2} \ \mbox{Re} \left( C(t,t') - \frac{{\rm i}\hbar}{2} R(t,t') \right)^{p-1}
\!\! ,
\\
\Sigma_e(\tau - \tau_1) &=& - \frac{J_0^2p}{2\hbar}  \; \left( q_d(\tau - \tau_1)  \right)^{p-1}
\!\! ,
\\
\Sigma^\lceil(t, \tau_1) &=& \frac{J_0Jp}{2\hbar}  \; \left( G^\lceil(t,\tau_1) \right)^{p-1}
\!\! ,
\\
\Sigma^\rceil(\tau_1, t) &=& \frac{J_0Jp}{2\hbar}  \; \left( G^\rceil(\tau_1,t)  \right)^{p-1}
\!\! .
 \end{eqnarray}
 The equation for $z_e$ is found from imposing $q_d(0)=1$~\cite{CuGrSa00,CuGrSa01}.

The boundary conditions on the  correlation function ensure the consistency between the equations above. For instance, a perfect matching is found by setting
 $\tau=0$ in Eq.~(\ref{eq:Qt1tau}) and $t'=0$ in Eq.~(\ref{eq:C}), 
 with the resulting equation determining the real-time
 evolution of the correlation with the initial time, {\it i.e.} $C(t,0)$.

We have chosen to present the equations using the self-energies $\Sigma$ and the 
vertex $D$ to show them in a generic writing that has, basically, the structure of 
the full set of equations in all other mean-field models.

\subsection{Classical limit}

One can readily check that in the classical limit
\begin{eqnarray}
q_d(\tau)&\to & C(0,0) \; =  \; 1 \quad \Rightarrow \quad \Sigma_e(\tau) \, \to \, -\frac{J_0^2p}{2\hbar}
\; ,
\\
G^\lceil(t,\tau) &\to &  C(t,0) \qquad \quad\Rightarrow \quad \Sigma^\lceil(t,\tau) \; \to \;  \frac{J_0 Jp}{2\hbar}  \left( C(t,0) \right)^{p-1}
\; ,
\\
\Sigma(t,t') &\to & \frac{J^2 p (p-1)}{2} C^{p-2}(t,t') R(t,t')
\; ,
\\
D(t,t') &\to & \frac{J^2 p}{2} C^{p-1}(t,t')
\; ,
\end{eqnarray}
and the equations of motion of the classical model derived and studied in~\cite{CuLoNe17}
are recovered. Indeed, the equations in which the free operator acts on the real time linear response $R$ and symmetric correlation
$C$ approach Eqs.~(50) and (51)  in Ref.~\cite{CuLoNe17} (for $\gamma=0$) with the
last term in the second equation reading
\begin{eqnarray}
{\footnotesize \frac{J J_0 p}{2\hbar}} \int_0^{\beta_0\hbar} d\tau_1 \ \left(G^\lceil(t,\tau_1)\right)^{p-1} {G^\lceil} (t',\tau_1)
&\mapsto&
\frac{J J_0 p}{2} \; \beta_0 \left(G^\lceil(t, 0)\right)^{p-1} {G^\lceil} (t',0)
\nonumber\\
&=& \frac{J J_0 p}{2} \; \beta_0 \;  C^{p-1}(t, 0)  C(t',0)
\; .
\end{eqnarray}
The equation in which the real time free operator acts on ${G^\lceil}(t_1,\tau)$
approaches Eq.~(51) in Ref.~\cite{CuLoNe17}
if we use ${G^\lceil}(t,\tau) \to C(t,0) $ and $q_d(0)=1$.
Finally, the classical limit of the equation that determines $q_d(\tau)$ is taken by first taking a Fourier transform and
rewriting this equation in term of $\tilde q(\omega_k)$, with $\omega_k$ the Matsubara frequencies, next making an analytic continuation
to real frequencies to relate $\tilde q$ to $R(\omega)$. 
The remaining equation is then the one on the classical linear response function~\cite{Mahan00}.

\subsection{Equilibrium}

We will study here the equations in a case in which $J_0=J$, that is to say with 
no quench. The system starts its dynamics at time $t=0$ in equilibrium and it
should remain in equilibrium ever after. Under such circumstances, 
all one real-time dependent quantities and, in particular, the Lagrange multiplier $z(t)$ 
should be constant. Moreover, all correlations and the 
linear response function should be stationary. 

The first equilibrium property we can check is the quantum fluctuation dissipation 
theorem (FDT) that relates the time-delayed linear response to the self correlation
function. This relation can be easily verified by comparing the first and second 
equations in the set. The argument goes as follows. First, one first notices that 
the last term Eq.~(\ref{eq:C}) is just a constant in equilibrium, 
since $\Sigma^\lceil$ and $G^\rceil$ cannot depend on $t$ and $t'$, respectively.
Then, the structure of the integrals, and the relations between the $\Sigma$ and $D$
kernels are such that the proof exhibited in~\cite{CuLo98,CuLo99} for the 
relaxational case can be simply applied to this problem as well. Therefore, the 
two real-time dependent equations (\ref{eq:R}) and (\ref{eq:C}) are consistent 
with the FDT relating $R$ and $C$ on the one hand, and $\Sigma$ and $D$ on the other.

Next, we prove that $z(t_1)=z=z_e$ in equilibrium.  To do it we work with the equations for ${G^\lceil}$ Eqs.~(\ref{eq:Qt1tau}) and one for $G^\rceil$ in which the derivative operator is written in the 
imaginary time domain (not shown here). 
We use the stationarity to change the two (real and imaginary) time 
arguments  to the combination
$t+i\tau$ and we then change, in one of the equations, the time-derivatives from
$\partial_{t^2}$ to $\partial_{\tau^2}$ (or {\it vice versa}). Having done this, we can 
compare the two equations and readily conclude that the two $z$'s should be the same,
$z=z_e$.

We now want to check that in the non-trivial region of the phase diagram between $(T_s,\Gamma_s)$ and $(T_d,\Gamma_d)$
\begin{eqnarray}
\lim_{t\to\infty} C(t,0) = \lim_{t-t'\to\infty} C(t,t')=q \neq 0
\end{eqnarray}
and that, moreover, $q$ is the value associated to one of the TAP states~\cite{BiCu01}.
Taking the limit $t\to\infty$ in the equation for $C(t,0)$ we find
\begin{equation}
z q = q \lim_{t\to\infty} \int_0^t dt_1 \; \Sigma(t,t_1) + q^{p-1} \frac{JJ_0p}{2\hbar} \int_0^{\beta\hbar} d\tau_1 \; q_d(\tau_1)
\label{eq:q0}
\end{equation}
with solution $q=0$ but also $q\neq 0$ in the above-mentioned region of the parameter space. 
This equation generalises to the quantum problem Eq.~(82) in
Ref.~\cite{CuLoNe17}. The argument presented in this paper to show 
that the non-trivial $q$ corresponds to the system being confined in a TAP 
state can also be adapted to prove the same result in the quantum problem.

The analysis of the equation for $C(t,t')$ in the limit $t-t'\to\infty$ follows the steps
in between Eqs.~(6.7) and (6.10) in Ref.~\cite{CuLo99}. The only differences are that there is no
aging contribution here. In the long times limit we have
\begin{eqnarray}
A_\infty \equiv \lim_{t\to\infty}\lim_{t'\to\infty} \frac{JJ_0p}{2\hbar} \int_0^{\beta\hbar} d\tau_1 \; {G^\lceil}^{p-1}(t,\tau_1) {G^\rceil}(\tau_1, t') =
 \frac{JJ_0p}{2} \beta q^p
 \; .
\end{eqnarray}
The condition on $q$ then reads
\begin{equation}
-zq+A_\infty + q \lim_{t\to\infty} \int_0^t dt_1 \; \Sigma(t-t_1) + \frac{J^2p}{2} q^{p-1} \lim_{t\to\infty} \int_0^t dt_1 \; R(t-t_1) =0
\; .
\label{eq:q}
\end{equation}
Comparing now Eqs.~(\ref{eq:q0}) and (\ref{eq:q}) now notices that they become the same equation if
\begin{equation}
\tilde R(\omega=0) = \lim_{t\to\infty} \int_0^t dt_1 \; R(t-t_1) = \frac{1}{\hbar} \int_0^{\beta\hbar} d\tau \; [q_d(\tau)-q]
\end{equation}
which is a statement of the FDT, a generic feature valid in equilibrium.

Therefore, we have shown that, for certain values of the parameters, 
the equations admit an equilibrium solution
with $q\neq 0$. This correspond to the
dynamics being confined to a TAP state, see~\cite{BiCu01} for more details on this approach.

\section{Conclusions}
\label{sec:conclusions}

The structure of the equations that we derived is very general and applies to more generic models and 
field theories with quenched
randomness, once studied on average over the frozen randomness, starting from canonical equilibrium conditions.

The full solution of the dynamics starting from all possible initial conditions in canonical equilibrium, after quenches of various
kind, can only be done numerically. This study will be presented elsewhere. We simply sketch here the steps that an algorithm
constructing the solution should follow:
\begin{enumerate}
\item[(i)]
Solve Eq.~(\ref{eq:qd}) and, in this way, find the value taken by $z_e$ and construct $q_d(\tau)$ for all $\tau$, as done in Refs.~\cite{CuGrSa00,CuGrSa01}
for this very same model.
\item[(ii)]
In the equations for the real time dependent linear response and symmetrised correlation functions, the purely imaginary time
the mixed $G^\lceil(t,\tau)$ and $G^\rceil(\tau,t)$ intervene (in the last term in the rhs of the equation for $C$) and by their means $q_d(\tau)$ does as well.
Therefore, we need to determine the  two mixed correlations simultaneously with $R$ and $C$ and this we do working
with the set of Eqs.~(\ref{eq:R}), (\ref{eq:C}), (\ref{eq:Qt1tau}) and (\ref{eq:Qtaut1bis}) at once.
\end{enumerate}
This is a hard task and we will present the results of the numerical investigation in a separate publication. A different approach to treat (paramagnetic-like) initial conditions in the same problem, without introducing the imaginary time branch of the contour is currently being pursued by 
Schir\`o {\it el al.}~\cite{Schiro}.

Path-integral formalisms as the one discussed here are useful not only to derive Schwinger-Dyson equations but also
to deduce more generic results, even out of equilibrium, such as the very much studied fluctuation theorems. In the 
quantum context, these theorems can be obtained exploiting a symmetry of the generating functional~\cite{SiChGaDi15,ArBiCu18}, and the correct treatment of thermal initial states~\cite{ArBiCu18} 
is fundamental to derive the searched generic relations. This is just one more reason why reaching a correct handling of generating functionals with special initial states~\cite{Gelis99,Sensarma18} 
is very important.

\vspace{0.5cm}

\noindent
{\bf Acknowledgements}
 This work was initiated many years ago together 
with D. R. Grempel and C. A. da Silva Santos. We are deeply indebted to them for this early collaboration. 
We thank M. Schir\'o, M. Tarzia and S. Thomson for very useful discussions.
We acknowledge financial support from ECOS-Sud A14E01, PICS 506691 (CNRS-
CONICET Argentina) and NSF under Grant No. PHY11-25915. LFC thanks the KITP Santa Barbara for hospitality during part of the
preparation of this work. She is a member of Institut Universitaire de
France.

\bibliographystyle{phaip}
\bibliography{notas}

\end{document}